\documentclass[aps,prd,twocolumn,nofootinbib]{revtex4-1}
\usepackage{epsfig}
\usepackage[colorlinks,linkcolor=blue,anchorcolor=blue,citecolor=blue,urlcolor=blue,breaklinks=true]{hyperref}
\usepackage{graphics}
\usepackage{color}
\usepackage{slashed}
\usepackage{makecell}
\usepackage{subfigure}
\usepackage{changes}
\usepackage{cancel}

\begin{document}
\author{Shu-Sheng Xu$^{1}$}~\email[]{Email: xuss@njupt.edu.cn}
\author{Chao Shi$^{2}$}~\email[]{Email:
cshi@nuaa.edu.cn}
\address{$^{1}$ School of Science, Nanjing University of Posts and Telecommunications (NJUPT), Nanjing 210023, China}
\address{$^{2}$ Department of Nuclear Science and Technology,
Nanjing University of Aeronautics and Astronautics, Nanjing 210016, China}

\title{Perturbative method to estimate meson masses in the framework of Bethe-Salpeter equation beyond its dominant interaction}
\begin{abstract}
We propose a novel method to calculate the meson mass in the framework of Dyson-Schwinger equation and Bethe-Salpeter equation, once their dominant interactions are identified. The method is based on the perturbation theory of matrix, which is widely used in quantum mechanics. 
Taking interactions other than the dominant ones as perturbations, we derive the first order correction of quark propagator. Implementing the perturbation on BSE, the mass correction at first order is then given. We exemplify this method with the well known Rainbow-Ladder (RL) truncation, and go beyond the RL using a simple model, i.e., the Munczek-Nemirovsky~(MN) model, by studying the pion and $\rho$ mesons mass shift. The results are all in good agreement with those obtained by fully solving the BSE beyond RL. Our perturbative method therefore can be used to give a semi-quantitative estimate of meson mass correction in cases when the BSEs are complicated by interactions that go beyond the dominant one.
\bigskip


\bigskip

\noindent PACS Number(s): 14.20.Dh, 11.10.St, 12.38.Lg, 13.40.Gp

\end{abstract}
\maketitle

\section{Introduction}\label{intro}

Bound state problem is important in QCD since all hadrons are composite particles constituted by elementary particles as quarks and gluons. The mass spectrum of the hadrons encode information of their sub-structure. Many approaches have been developed to study hadron spectrum, such as constituent quark model~\cite{amsler2004mesons,roberts2008heavy,vijande2005constituent,PhysRevD.67.014027,ebert2010heavy,PhysRevD.81.034010}, Nambu-Jona-Lasinio~(NJL) model~\cite{alkofer1990pseudoscalar,clemens2017holographic}, functional renormalization group~\cite{PhysRevD.99.054029,PhysRevD.97.054006}, lattice QCD~\cite{karsch2003hadron,PhysRevD.77.034501,PhysRevLett.103.262001,dougall2003spectrum,PhysRevD.82.014507,PhysRevLett.105.252002} and Bethe-Salpeter equation (BSE)~\cite{watson2004bethe, PhysRevD.83.096006, PhysRevD.59.094001, PhysRevC.56.3369,PhysRevC.75.045201,PhysRevC.67.035202,PhysRevC.60.055214,PhysRevC.70.042203,PhysRevC.85.035202}. Among them, the BSE coupled with Dyson-Schwinger equation~(DSE) provides an efficient tool. The BSE-DSE approach starts with the quark and gluon degrees of freedom, and preserves the symmetries of QCD, such as the $U(1)$ gauge symmetry and $U_A(1)$ chiral symmetry~\cite{PhysRevD.91.054035}. In practice, this is realized by implementing truncations (on vertices and interaction kernels) that  respect the vector Ward-Takahashi identity~(WTI) and axial-vector WTI~\cite{PhysRevD.52.4736,MARIS1998267}.  During the last two decades, the rainbow-ladder~(RL) truncation achieved great success in describing $J^{P}=0^-, 1^-$ ground state mesons~\cite{MARIS1998267,WILLIAMS2019134943,PhysRevC.60.055214,PhysRevD.93.034026,PhysRevD.93.074010} and $\frac{1}{2}^+, \frac{3}{2}^+$ ground state baryons~\cite{PhysRevLett.104.201601,PhysRevD.84.096003}, which are regarded as orbital angular momentum $L=0$ dominant states. However, in the study of radially excited pseudo-scalar and vector mesons, the RL truncation generally underestimates their masses, implying the necessity of  going beyond RL truncation~\cite{krassnigg2008excited}.

It is a subtle work to construct dressed quark-gluon vertex and quark-anti-quark interaction kernel beyond RL truncation. There are mainly two approaches. One is to use explicit diagrammatic representation to the DSE of the dressed quark-gluon vertex~\cite{BENDER19967,watson2004bethe,PhysRevC.70.035205,PhysRevC.75.045201,ALKOFER2009106,PhysRevD.76.094009,PhysRevLett.103.122001,PhysRevD.78.074006,PhysRevD.92.054030}.
In Ref.~\cite{BENDER19967}, the first model study of meson BSEs beyond RL truncation is given by considering sub-leading Abelian correction for quark-gluon vertex and quark-anti-quark kernel, followed by a number of further studies \cite{PhysRevC.65.065203,watson2004bethe,PhysRevC.70.035205,PhysRevC.75.045201}. The non-Abelian quark-gluon vertex is discussed in Ref.~\cite{ALKOFER2009106}, and the meson spectrum are calculated~\cite{PhysRevLett.103.122001}. Pion exchange between quarks  was also considered~\cite{PhysRevD.78.074006,williams2010bethe}.
Another approach to go beyond RL truncation is to construct the tensor structures of the dressed quark-gluon vertex and thereafter to build the quark-anti-quark interaction kernel under the constrain of axial-vector WTI~\cite{PhysRevD.72.094025,PhysRevLett.106.072001,PhysRevC.85.052201,PhysRevLett.103.081601,PhysRevD.93.096010}. 

In this paper, we propose a new method to estimate the meson mass correction, once the dominant interaction is identified. This includes the case of RL truncation, given the success of RL  truncation in describing the properties of ground state hadrons, e.g., the masses and the decay constants of pion and $\rho$ mesons~\cite{PhysRevC.56.3369,PhysRevC.60.055214,PhysRevC.84.042202}. Our starting point is to treat the interaction terms beyond the dominant term as perturbations, i.e., their contribution to the DSE and BSE are small as compared to the dominant term. As compared to fully solving the DSE-BSE, our method requires much less computational effort. Furthermore, as we will show below, it serves as a general method which can be conveniently applied to any interaction kernel. 

This paper is organized as follows. In Sec.~\ref{dsebse}, we give a brief introduction to DSE-BSE approach. In Sec.~\ref{method}, we introduce our new perturbative method. The formula of quark propagator and the meson mass correction at the first order is derived in detail.
In Sec.~\ref{application}, we employ a simple model, the Munczek-Nemirovsky (MN) model, and revisit the pion and $\rho$ meson study with our new method. In Sec.~\ref{sum}, we summarize this work.
\section{Dyson-Schwinger equation and Bethe-Salpeter equation}\label{dsebse}
\begin{figure}[!h]
\centering
\includegraphics[width=0.4\textwidth]{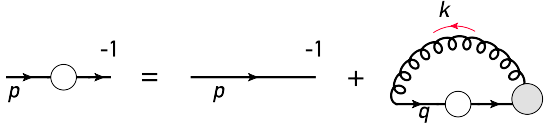}
\caption{The graphical representation of the quark Dyson-Schwinger equation.} \label{fig1}
\end{figure}

The elementary degrees of freedom in QCD are quarks and gluons.
The quark DSE describes the equation of motion of the quark propagator, revealing how quark propagator is determined by its interaction with gluons. For a given flavor of quark, the quark DSE, as diagrammatically represented in Fig.~\ref{fig1}, is
\begin{eqnarray}
S^{-1}(p) &=& Z_2(i\gamma\cdot p + Z_m m) + \Sigma(p),  \label{quarkdse}
\\
\Sigma(p) &=& Z_1\int_q g^2D_{\mu\nu}(k) \gamma_\mu\frac{\lambda^a}{2} S(q) \Gamma^a_\nu(p,q), \label{selfenergy}
\end{eqnarray}
where $S(p)$ is the dressed quark propagator. The $\int_q$ is the abbreviation of $\int\frac{d^4q}{(2\pi)^4}$. Here the $m$ is the current quark mass, and $\Sigma(p)$ is the quark self-energy. $D_{\mu\nu}(k)$ is the dressed gluon propagator with $k=p-q$, $\lambda^a, (a=1\cdots 8)$ are Gell-Mann matrices, and $\Gamma^a_\nu(p,q)$ is the dressed quark-gluon vertex.  The $Z_1$, $Z_2$ and $Z_m$ are the renormalization constants of the quark-gluon vertex, quark wave-function and quark mass respectively. The dressed gluon propagator and the dressed quark-gluon vertex satisfy their own DSEs, which are related to higher-point Green functions. Therefore the quark DSE is not closed. In practical study,  truncations and ansatzs for effective interaction must be employed so that the quark DSE gets closed and solvable.

The dressed quark propagator has the general structure
\begin{equation}
S(p) = -i\gamma\cdot p\sigma_V(p^2) + \sigma_S(p^2),
\end{equation}
and the structure of the inverse of the dressed quark propagator is
\begin{equation}
S^{-1}(p) = i\gamma\cdot p A(p^2) + B(p^2).
\end{equation}
The scalar functions are related by
\begin{eqnarray}
\sigma_V(p^2) &=& \frac{A(p^2)}{p^2A^2(p^2)+B^2(p^2)},  \label{relations1}
\\
\sigma_S(p^2) &=& \frac{B(p^2)}{p^2A^2(p^2)+B^2(p^2)}.  \label{relations2}
\end{eqnarray}
\begin{figure}[!h]
\centering
\includegraphics[width=0.35\textwidth]{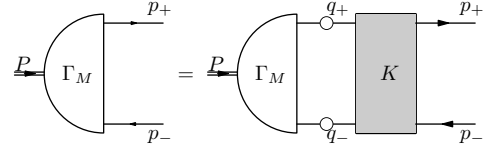}
\caption{The graphical representation of the meson Bethe-Salpeter equation.} \label{fig2}
\end{figure}

On the other hand, the quark-antiquark two-body bound state is governed by the Bethe-Salpeter equation, as diagrammatically represented in Fig.~\ref{fig2},
\begin{eqnarray}
\Gamma_M(p,P) &=& \int_q K(p,q,P) \chi_M(q,P),  \label{mesonbse}
\\
\chi_M(q,P) &=& S(q_+)\Gamma_M(q,P)S(q_-),
\end{eqnarray}
where $\Gamma_M(p,P)$ is the Bethe-Salpeter amplitude~(BSA) of meson. The $\chi_M$ is the meson wave function, $q_\pm=q\pm\eta_\pm P$ with $\eta_++\eta_-=1$, and $K(p,q,P)$ is the interaction kernel.

The general structure of Bethe-Salpeter amplitude for different $J^{P}$ meson is different. Take pseudo-scalar meson as an example, i.e. $J^{P}=0^{-}$, the most general amplitude reads
\begin{equation}
\Gamma_{0^-}(p,P) = \sum_{i=1}^4 T_{0^-}^i(p,P) F^i_{0^-}(p^2,p\cdot P),
\end{equation}
with
\begin{equation}
T_{0^-}^i=\left\{\gamma_5,\quad \gamma_5\gamma\cdot P,\quad \gamma_5\gamma\cdot p,\quad [\gamma\cdot p,\gamma\cdot P]\right\},
\end{equation}
and $F^i_{0^-}(p^2,p\cdot P)$ are scalar functions. Inserting the general BSA into the BSE, which is a homogeneous equation. The determination of meson mass can further be transformed into an eigenvalue problem, e.g.,
\begin{equation}
\lambda(P^2) \Gamma_M(p,P) = \int_q K(p,q,P)\chi_M(q,P).  \label{mesonbse2}
\end{equation}
The calculated meson mass is located at $m_M$ with $\lambda(-m_M^2)=1$.

 The BSE contains on one hand the dressed quark propagator, which is the solution to quark DSE Eq.~(\ref{quarkdse}), and on the other hand the kernel should be constructed together with the dressed quark-gluon vertex, i.e., they are constrained by vector and axial-vector WTIs,
\begin{eqnarray}
iP_\mu \Gamma^\gamma_\mu(k,P) &=& S^{-1}(k_+) - S^{-1}(k_-),
\\
P_\mu\Gamma_{5\mu}(k,P) + 2im\Gamma_5(k,P) &=& S^{-1}(k_+)i\gamma_5 - i\gamma_5S^{-1}(k_-).
\nonumber\\
\end{eqnarray}
Here $\Gamma^\gamma_\mu(k,P)$ is the photon-quark vertex.
The vector WTI guarantees the gauge symmetry $U(1)$, and the axial-vector WTI guarantees the chiral symmetry of the QCD. Combined with the inhomogeneous BSEs of the vector and axial-vector vertex and the quark gap equation, one can relate the kernel to the dressed quark-gluon vertex as~\cite{qin2016systematic}
\begin{eqnarray}
&&\int_q K_{\alpha\alpha^\prime,\beta^\prime\beta} \left[ S(q_-) - S(q_+) \right]_{\alpha^\prime\beta^\prime}
\nonumber\\
&=& \int_q D_{\mu\nu}(k-q)\gamma_\mu \Big[ S(q_+)\Gamma_\nu(q_+,k_+)
\nonumber\\
&&\hspace*{25mm} - S(q_-)\Gamma_\nu(q_-,k_-) \Big],
\label{kernelvertex1}
\end{eqnarray}
\begin{eqnarray}
&&\int_q K_{\alpha\alpha^\prime,\beta^\prime\beta}\left[ S(q_+)\gamma_5 + \gamma_5 S(q_-) \right]_{\alpha^\prime\beta^\prime}
\nonumber\\
&=& \int_q D_{\mu\nu}(k-q) \gamma_\mu \Big[ S(q_+)\Gamma_\nu(q_+,k_+)\gamma_5
\nonumber\\
&&\hspace*{25mm} - \gamma_5S(q_-)\Gamma_\nu(q_-,k_-) \Big].
\label{kernelvertex2}
\end{eqnarray}
The above equations must be satisfied in constructing the kernel for a specified dressed quark-gluon vertex and vice versa.

\section{Go beyond the dominant interaction perturbatively in DSE and BSE.} \label{method}
Without loss of generality, one can write the dressed quark-gluon vertex and quark-anti-quark interaction kernel as
\begin{eqnarray}
g^2D_{\mu\nu}(k)\Gamma_\nu(p,q) = \mathcal{G}(k^2)D^{\mathrm{free}}_{\mu\nu}(k) \left(\Gamma^0_\nu + \epsilon\Gamma^I_\nu(p,q) \right),
\nonumber\\
\label{fullvertex}
\end{eqnarray}
and
\begin{equation}
K_{\alpha\alpha^\prime,\beta^\prime\beta}(p,q,P) = \mathcal{G}(k)D^{\mathrm{free}}_{\mu\nu}(k)\left( {K^0}^{\mu\nu}_{\alpha\alpha^\prime,\beta^\prime\beta} + \epsilon{K^I}^{\mu\nu}_{\alpha\alpha^\prime,\beta^\prime\beta}\right).   \label{kernelmodel2}
\label{fullkernel}
\end{equation}
where $D^{\mathrm{free}}_{\mu\nu}(k)=\delta_{\mu\nu} -(1-\xi) \frac{k_\mu k_\nu}{k^2}$ is the free gluon propagator. The $\xi$ is the gauge parameter and we use Landau gauge $\xi=0$ in this work. $\mathcal{G}(k)$ is the effective interaction in both Eqs.~(\ref{fullvertex}) and (\ref{fullkernel}), which absorbs the coupling $g^2$, the dressing function of gluon propagator and some momentum dependence of the dressed quark-gluon vertex. We then assume the $\Gamma^0_\nu(p,q)$ and ${K^0}^{\mu\nu}_{\alpha\alpha^\prime,\beta^\prime\beta}$ are the dominant part of quark-gluon vertex and the quark-antiquark kernel, the $\Gamma^I_\nu(p,q)$ and ${K^I}^{\mu\nu}_{\alpha\alpha^\prime,\beta^\prime\beta}$ can be regarded as perturbation, denoted by the small expansion parameter $\epsilon$. 

\subsection{quark DSE}
Denoting $S_0(p)$ as the solution to quark DSE within dominant truncation, the quark propagator with the full quark-gluon vertex $S(p)$ can be written as
\begin{equation}
S(p) = S_0(p) + \epsilon S_1(p) + \mathcal{O}(\epsilon^2),  \label{quarkpert1}
\end{equation}
or equivalently,
\begin{equation}
S^{-1}(p) = S_0^{-1}(p) + \epsilon S_1^{-1}(p) + \mathcal{O}(\epsilon^2). \label{quarkpert2}
\end{equation}
The scalar functions take the expansion analogously
\begin{eqnarray}
\sigma_V(p^2) &=& \sigma_{V0}(p^2) + \epsilon \sigma_{V1}(p^2) + \mathcal{O}(\epsilon^2), \label{quarkpert3}
\\
\sigma_S(p^2) &=& \sigma_{S0}(p^2) + \epsilon \sigma_{S1}(p^2) + \mathcal{O}(\epsilon^2), \label{quarkpert4}
\\
A(p^2) &=& A_0(p^2) + \epsilon A_1(p^2) + \mathcal{O}(\epsilon^2), \label{quarkpert5}
\\
B(p^2) &=& B_0(p^2) + \epsilon B_1(p^2) + \mathcal{O}(\epsilon^2). \label{quarkpert6}
\end{eqnarray}
Based on their relations, i.e., Eqs.~(\ref{relations1}) and (\ref{relations2}), one obtains
\begin{eqnarray}
\sigma_{V1}(p^2) &=& \frac{B_0^2 A_1 - 2A_0 B_0 B_1 - p^2A_0^2 A_1}{\left(p^2A_0^2+B_0^2\right)^2}, \label{quarkpert7}
\\
\sigma_{S1}(p^2) &=& \frac{p^2 A_0^2 B_1 - 2p^2A_0 B_0 A_1 - B_0^2 B_1}{\left(p^2A_0^2+B_0^2\right)^2}. \label{quarkpert8}
\end{eqnarray}
Analogously, the quark DSE can also be expanded in $\epsilon$, which reads at the first order,
\begin{equation}
S_0^{-1}(p) + \epsilon S_1^{-1}(p) = iZ_2\gamma\cdot p + Z_4 m + \Sigma_0(p) + \epsilon\Sigma_1(p),
\end{equation}
the $\Sigma_0(p)$ is the quark self-energy Eq.~(\ref{selfenergy}) in dominant truncation, and $\Sigma_1(p)$ is
\begin{equation}
\Sigma_1(p) = \int_q \mathcal{G}(k^2)D^{\mathrm{free}}_{\mu\nu}(k)\gamma_\mu \left( S_0(q)\Gamma^1_\nu(p,q) + S_1(q)\gamma_\nu\right).
\end{equation}
The renormalization constants $Z_2, Z_4$ are
\begin{eqnarray}
Z_2 &=& Z_{20} + \epsilon Z_{21} + \mathcal{O}(\epsilon^2),
\\
Z_4 &=& Z_{40} + \epsilon Z_{41} + \mathcal{O}(\epsilon^2),
\end{eqnarray}
with
\begin{eqnarray}
Z_{21} &=& - \frac{1}{3p^2} \int_q \mathcal{G}(k^2) D_{\mu\nu}^{\mathrm{free}}(k) \mathrm{Tr}\Big[ \slashed{p}\gamma_\mu S_0(q)\Gamma^I_\nu(p,q)
\nonumber\\
&&\hspace*{20mm} +\slashed{p}\gamma_\mu S_1(q) \Gamma^0_\nu(p,q) \Big]\Big|_{p^2=\mu^2}.
\\
Z_{41} &=& - \frac{1}{3m}\int_q \mathcal{G}(k^2) D_{\mu\nu}^{\mathrm{free}}(k) \mathrm{Tr}\Big[ \gamma_\mu S_0(q)\Gamma^I_\nu(p,q)
\nonumber\\
&&\hspace*{20mm} + \gamma_\mu S_1(q) \Gamma^0_\nu(p,q) \Big]\Big|_{p^2=\mu^2}.
\end{eqnarray}
As the first order expansion in $\epsilon$, one has,
\begin{eqnarray}
S^{-1}_1(p) = iZ_{21}\gamma\cdot p + Z_{41}m + \Sigma_1(p).
\label{pertquark}
\end{eqnarray}
The above equation can be converted into coupled equations of $A_1(p^2)$ and $B_1(p^2)$, a unique solution can be found because of its linearity.
\subsection{The meson BSE}
We now turn to the BSE, Eq.~(\ref{mesonbse2}). The BS amplitude $\Gamma_M(p,P)$ and the eigenvalue $\lambda(P^2)$ of Eq.~(\ref{mesonbse2}) depend on the small parameter $\epsilon$, so they be expanded in the same way as the quark propagator,
\begin{eqnarray}
\Gamma_M(p,P) &=& \Gamma_{M0}(p,P) + \epsilon \Gamma_{M1}(p,P) + \mathcal{O}(\epsilon^2), \label{bsapert}
\\
\lambda(P^2) &=& \lambda_0(P^2) + \epsilon\lambda_1(P^2) + \mathcal{O}(\epsilon^2). \label{lambdapert}
\end{eqnarray}
Here the $\Gamma_{M0}(p,P)$ and $ \lambda_0(P^2) $ are the results within the dominant truncation. Inserting Eqs.~(\ref{quarkpert1}), (\ref{bsapert}) and (\ref{lambdapert}) into Eq.~(\ref{mesonbse2}), one obtains
\begin{eqnarray}
&&\left(\lambda_0+\epsilon\lambda_1\right)\left(\Gamma_{M0}+\epsilon\Gamma_{M1}\right)
\nonumber\\
&=& \int_q\left(K_0+\epsilon K^I\right) \left(S^+_0+\epsilon S^+_1\right) \left(\Gamma_{M0}+\epsilon \Gamma_{M1}\right) \left(S^-_0+\epsilon S^-_1\right),
\nonumber\\
\end{eqnarray}
where $S^\pm_{0,1}=S_{0,1}(q\pm P/2)$.

With the help of the zeroth order of BSE, the first ordered BSE reads
\begin{eqnarray}
&&\lambda_0\Gamma_{M1} - \int_q K_0 S^+_0\Gamma_{M1}S^-_0
\nonumber\\
&=& \int_q\left( K^IS^+_0\Gamma_{M0}S^-_0 + K_0S^+_1\Gamma_{M0}S^-_0 + K_0S^+_0\Gamma_{M0}S^-_1 \right)
\nonumber\\
&& - \lambda_1\Gamma_{M0}.
\nonumber\\
\label{bsepert}
\end{eqnarray}
Multiply $S^-_0\overline{\Gamma}_{M0}S^+_0$ on the left and take trace for both sides of Eq.~(\ref{bsepert}), the left-hand-side vanishes with the help of conjugated BSE. The first order perturbation of eigenvalue is
\begin{eqnarray}
\lambda_1 &=& \frac{1}{\mathcal{N}_0}\mathrm{Tr}\bigg[\int_\ell S^-_0\overline{\Gamma}_{M0}S^+_0 \int_q\Big( K^I S^+_0\Gamma_{M0}S^-_0
\label{lambda1}
\nonumber\\
&&\hspace*{10mm}+ K_0S^+_1\Gamma_{M0}S^-_0 + K_0S^+_0\Gamma_{M0}S^-_1 \Big)\bigg].
\\
\mathcal{N}_0 &=& \mathrm{Tr}\int_\ell S^-_0\overline{\Gamma}_{M0}S^+_0\Gamma_{M0}
\label{n0}
\end{eqnarray}
All these elements are known from zeroth quark DSE, meson BSE and the first order quark DSE. One can search the meson mass $m_M$ so that $\lambda(P^2=-m_M^2)=\lambda_0+\epsilon\lambda_1=1$.

However, a critical remedy of Eq.~(\ref{lambda1}) is needed, if we consider the case of pion. As we know, the pion is the Goldstone boson of chiral symmetry. It is massless in the exact chiral limit $m=0$. But the expansion Eqs.~(\ref{bsapert}) and (\ref{lambdapert}) can not automatically preserve this property. Additional constraints should be taken into consideration. In the chiral limit, the eigenvalue of pion BSE satisfies the same expansion as Eq.~(\ref{lambdapert}), and $\lambda^{\pi CL}(P^2=0)=\lambda_0^{\pi CL}(P^2=0)=1$ because of the dominant truncation and full interaction both preserve chiral symmetry, the superscript ``CL" refers to ``chiral limit''. Hence
\begin{equation}
0=\sum_{i=1}^{\infty} \epsilon^i \lambda_i^{\pi CL}(P^2=0). \label{chiralrelation}
\end{equation}
Subtracting Eq.~(\ref{chiralrelation}) from Eq.~(\ref{lambdapert}), the eigenvalue of meson BSE can be expanded as
\begin{equation}
\lambda(P^2) = \lambda_0(P^2) + \epsilon \lambda^R_1(P^2)  + \mathcal{O}(\epsilon^2),
\end{equation} 
with
\begin{equation}
\lambda^R_1(P^2) = \lambda_1(P^2) - \lambda_{1}^{\pi CL}(0).
\end{equation}
This is our modified (and final) result concerning $\lambda$.
It is obvious that the pion is massless at every order of $\epsilon$ in the chiral limit. Both of the $\lambda_1(P^2)$ and $\lambda_1^{\pi CL}(0)$ can be calculated by Eqs.~(\ref{lambda1}) and (\ref{n0}). 

\section{Go beyond the RL truncation perturbatively in DSE and BSE.}\label{application}
\begin{figure}[htb]
\centering
\includegraphics[width=0.35\textwidth]{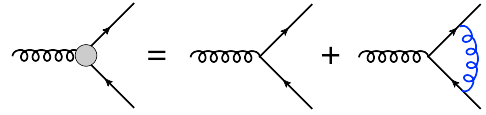}
\caption{The quark gluon-vertex on the next leading order.} \label{quarkgluonvertex}
\end{figure}

\begin{figure}[htb]
\centering
\includegraphics[width=0.35\textwidth]{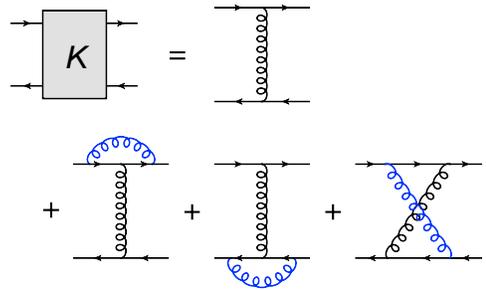}
\caption{The quark-anti-quark interaction kernel on the next leading order.} \label{quarkkernel}
\end{figure}

In this section we exemplify our perturbative approach with a  specific calculation of the pion and $\rho$ meson by going beyond the RL truncation. We consider the quark-gluon vertex 
\begin{equation}
\Gamma_\nu(p,q) = \gamma_\nu + \frac{1}{6} \int_\ell g^2 D_{\rho\sigma}(\ell)\gamma_\rho S(p-\ell)\gamma_\nu S(q-\ell)\gamma_\sigma,
\label{dressedvertex}
\end{equation}
which is also diagrammatic represented as in Fig.~\ref{quarkgluonvertex}.
The corresponding quark-anti-quark interaction kernel is shown in Fig.~\ref{quarkkernel}.

Apprarently, the first terms on the right hand side of Fig.~\ref{quarkgluonvertex} and Fig.~\ref{quarkkernel} constitute the RL truncation. It is known that the RL truncation dominates in the case of pion and $\rho$,  so we take it as the leading term, with the rest terms as perturbations.

In this section, we employ the Munczek-Nemirovsky~(MN) model~\cite{PhysRevD.28.181} to illustrate how to use the perturbative method, and analysis the numerical results. The MN model for the effective interaction is
\begin{equation}
\label{eq:Gk}
\mathcal{G}(k) = G(2\pi)^4\delta^{4}(k),
\end{equation}
where $G=0.281~\mathrm{GeV}^2$ in this work, and we use current quark mass $m=0.012$~GeV~\cite{BENDER19967}. It has the dynamical mass generation effect built in by imposing a strong interaction strength. In terms of ground $\pi$ and $\rho$ masses, the availability of MN model has been tested~\cite{BENDER19967}. However, it is worth to stress that the MN model is a much simplified model. The delta function implies zero momentum exchange between quarks, which precludes explicit gauge sector interactions, such as the three-gluon vertex effects. In mesons where non-Abelian interactions are important, finding the appropriate dominant contribution becomes an important task.

\begin{figure}[tb]
\centering
\includegraphics[width=0.22\textwidth]{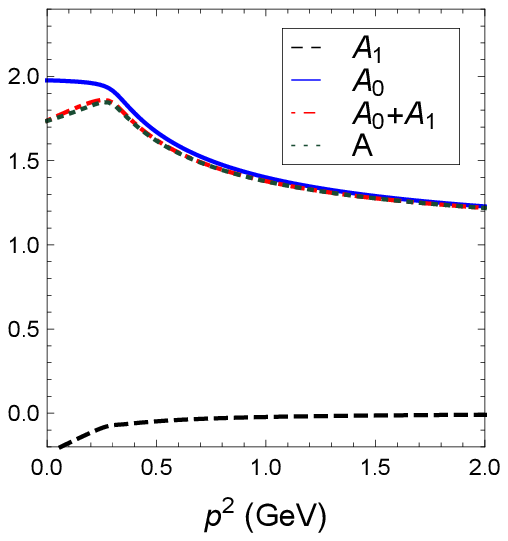}  \label{figA}
\includegraphics[width=0.22\textwidth]{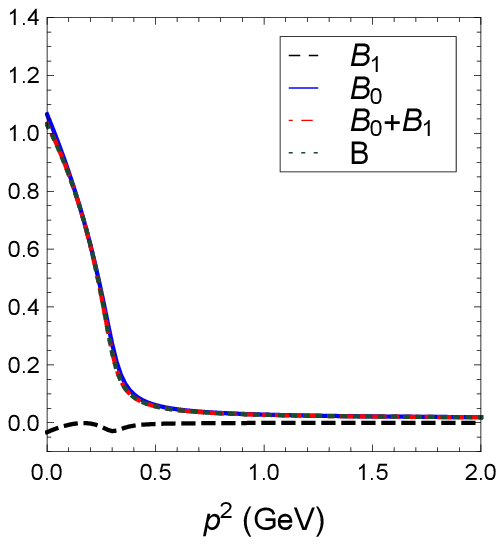}  \label{figB}
\caption{The scalar functions of quark propagator as function of $p^2$ in three cases: rainbow truncation $A_0~(B_0)$, perturbative results upto the first order, $A_0+A_1~(B_0+B_1)$ and non-perturbative results $A~(B)$.}
\label{figAB}
\end{figure}

Given Eq.~(\ref{eq:Gk}), the zeroth ordered quark DSE then reads
\begin{equation}
S_0^{-1}(p) = i\gamma\cdot p + m + G\gamma_\mu S_0(p) \gamma_\mu,
\end{equation}
Note the renormalization constants $Z_2=1$ and $Z_m=1$ because of the effective interaction is strongly suppressed in the ultraviolet.
One can derive the coupled equations for $A_0(p^2)$ and $B_0(p^2)$ as
\begin{eqnarray}
A_0(p^2) &=& 1 + 2G \frac{A_0(p^2)}{p^2A_0^2(p^2)+B_0^2(p^2)},
\\
B_0(p^2) &=& m + 4G \frac{B_0(p^2)}{p^2A_0^2(p^2)+B_0^2(p^2)}.
\end{eqnarray}
They have several unphysical solutions, e.g.,  the mass function of some solutions are negative. The physical solution is displayed as $A_0$, $B_0$ in Fig.~\ref{figAB}, for which we constrain the mass function to be positive definite and $A_0(\infty)=1$, $B_0(\infty)=m$. The $B_0$ function has an evident rapid enhancement in the infrared region, as well as for the mass function $M_0(p^2)=B_0(p^2)/A_0(p^2)$, which is a clear sign of the dynamical chiral symmetry breaking.

The quark-anti-quark interaction kernel is illustrated in Fig.~\ref{quarkkernel}: the first term is the ladder approximation, and the second line is regarded as the first order in $\epsilon$.
The homogeneous BSE of meson in the ladder approximation with MN model is
\begin{equation}
\Gamma_{M0}(p,P) = -G\gamma_\mu S_0(p_+) \Gamma_{M0}(p,P) S_0(p_-)\gamma_\mu. \label{mnbse}
\end{equation}
The $\delta$ function in the effective interaction entails that the bound state have zero relative momentum, so the BSA is relative momentum independent.

For numerical convenience, we introduce projector of Dirac-Lorentz structures, $\overline{T}^i_{J^P}(P)$, so that
\begin{equation}
\mathrm{Tr}\left[ \overline{T}^i_{J^P}(P) T^j_{J^P}(P) \right] = \delta_{ij}.
\end{equation}
For $J^P=0^-$ meson,
\begin{eqnarray}
T^1_{0^-} &=& i\gamma_5,\hspace*{10mm} T^2_{0^-} = \slashed{P}\gamma_5,
\\
\overline{T}^1_{0^-} &=& \frac{-i}{4}\gamma_5,\hspace*{6.5mm} \overline{T}^2_{0^-} = -\frac{1}{4P^2}\slashed{P}\gamma_5,
\end{eqnarray}
and for $J^P=1^-$ meson,
\begin{eqnarray}
T^{1\mu}_{1^-} &=& \gamma_\mu - \frac{\slashed{P}P_\mu}{P^2},\hspace*{12mm}  T^{2\mu}_{1^-} = \sigma_{\mu\nu}P_\nu.
\\
\overline{T}^{1\mu}_{1^-} &=& \frac{1}{12}\left(\gamma_\mu - \frac{\slashed{P}P_\mu}{P^2} \right),\hspace*{3mm} \overline{T}^{2\mu}_{1^-} = \frac{1}{12P^2}\sigma_{\mu\nu}P_\nu.
\end{eqnarray}
The general structure of $\pi$ and $\rho$ meson are
\begin{equation}
\Gamma_{\pi 0}(P) = \sum_{i=1}^2 T^i_{0^-}F^1_{\pi 0}(P^2), \label{pionbsa}
\end{equation}
for pion, and
\begin{equation}
\Gamma_{\rho 0}^\mu(P) = \sum_{i=1}^2 T^{i\mu}_{1^-}F^i_{\rho 0}(P^2), \label{rhobsa}
\end{equation}
for $\rho$ meson.

Inserting Eq.~(\ref{pionbsa}) into Eq.~(\ref{mnbse}), multiply $\overline{T}^i_{0^-}$ and take trace on both sides, one can write the pion BSE as an eigenvalue equation,
\begin{equation}
\lambda_0(P^2)\left(
\begin{array}{cc}
F^1_{\pi 0}  \\
F^2_{\pi 0}
\end{array}
\right) = 
\left(
\begin{array}{cc}
\mathcal{K}^\pi_{011}  &\mathcal{K}^\pi_{012} \\
\mathcal{K}^\pi_{021}  &\mathcal{K}^\pi_{022}
\end{array}
\right)
\left(
\begin{array}{cc}
F^1_{\pi 0} \\
F^2_{\pi 0}
\end{array}
\right),
\end{equation}
with
\begin{eqnarray}
\mathcal{K}^\pi_{011} &=& G\left( 4\sigma_{S0}^+\sigma_{S0}^- - P^2\sigma_{V0}^+\sigma_{V0}^- \right),
\\
\mathcal{K}^\pi_{012} &=& -2G P^2\left( \sigma_{V0}^+\sigma_{S0}^- + \sigma_{V0}^-\sigma_{S0}^+ \right),
\\
\mathcal{K}^\pi_{021} &=& -G \left( \sigma_{V0}^+\sigma_{S0}^- + \sigma_{V0}^-\sigma_{S0}^+ \right),
\\
\mathcal{K}^\pi_{022} &=& G \left( P^2\sigma_{V0}^+\sigma_{V0}^- -2 \sigma_{S0}^+\sigma_{S0}^- \right),
\end{eqnarray}
where $\sigma_{V0,S0}^\pm = \sigma_{V0,S0}(P^2/4)$.

\begin{figure}[htb]
\centering
\subfigure[The eigenvalues of $\pi$ meson in three cases.]{
\includegraphics[width=0.23\textwidth]{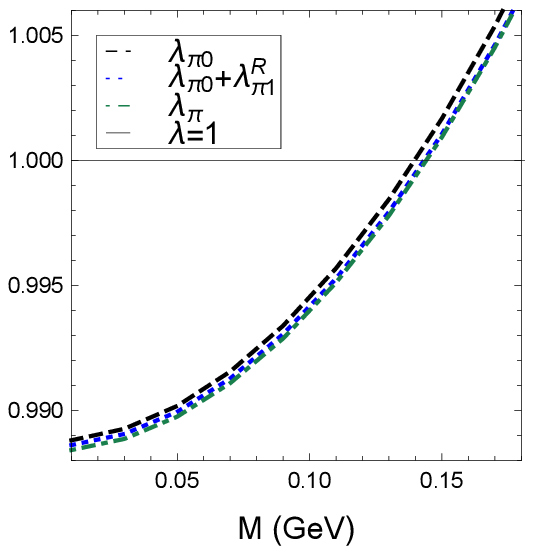}  \label{figlambda1}
}
\subfigure[The eigenvalues of $\rho$ meson in three cases.]{
\includegraphics[width=0.22\textwidth]{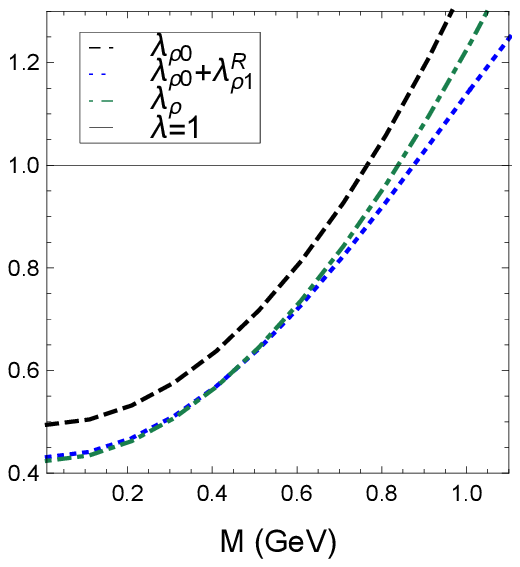}  \label{figlambda2}
}
\caption{The eigenvalues in three cases: $\lambda_{\pi 0,\rho 0}$ is the zeroth order case, $\lambda_{\pi 0,\rho 0}+\lambda_{\pi 1,\rho 1}^R$ is the correction upto the first order case and $\lambda_{\pi,\rho}$ is the non-perturbative results beyond RL truncation.}
\label{figpionrho}
\end{figure}

The eigenvalue and eigenvector of $\mathcal{K}_0^\pi$ can be calculated straightforwardly. The eigenvalues of $\pi$ and $\rho$ BSEs varying with $M=\sqrt{-P^2}$ are displayed as $\lambda_{\pi 0}$ and $\lambda_{\rho 0}$ in Fig.~\ref{figpionrho}. We obtain $M_{\pi 0}=0.140$ GeV and $M_{\rho 0}=0.767$ GeV.

Using the solutions of leading order, we calculate the first order of quark propagator $S_1$ with Eq.~(\ref{pertquark}), and display the numerical results in Fig.~\ref{figAB}. We can see from Fig.~\ref{figAB} that the first order correction for quark propagator is small as compared to $A_0$ and $B_0$. We remind this justifies our assumption on the beyond RL truncation term as a perturbation. To check our calculation further, we also calculated the full result, which is denoted by unlabeled $A$ and $B$ functions. These full results are obtained by aligning the DS Eq.~(\ref{quarkdse}) and BS Eq.~(\ref{mesonbse}) with full vertex Eq.~(\ref{dressedvertex}) and kernel $K$ displayed in Fig.~\ref{quarkkernel}. This is usually computational expensive but much simplified with MN model. From Fig.~\ref{figAB}, we see our perturbative technique gives results close to the full results, i.e., $A_0+A_1\approx A$ and $B_0+B_1\approx B$.

We further calculate the first order correction of eigenvalue for both $\pi$ and $\rho$ mesons using Eq.~(\ref{lambda1}), which consist of two terms. Given the $S_0$, $S_1$ and $\Gamma_{M0}$ we have, the calculation is straightforward. In Fig.~\ref{figpionrho}, we display eigenvalues for three cases, i.e., the RL results $\lambda_{M0}, (M=\pi,\rho)$, the perturbative results up to the first order $\lambda_{M0} +\lambda_{M1}^R$, and the full results $\lambda_M$. Again, the $\lambda_M$ is obtained by fully solving the DSE and BSE with the full interaction. In the Fig.~\ref{figlambda1}, we can see that the pion obtains a positive mass correction, the perturbative result  is highly quantitatively coincidence with the full result. The results of the $\rho$ meson can be seen in Fig.~\ref{figlambda2}. The first order correction is positive, which agrees with the full result semi-quantitatively. To conclude, we find that our perturbative method gives a semi-qualitatively consistent correction for the meson mass beyond RL truncation. It is convenient to use. For interaction kernels that are too complicated to compute, this may provide a first possible estimate over the mass shift beyond  dominant truncation.
\section{Summary}\label{sum}
We propose a novel method to calculate the mass correction beyond  dominant truncation in the framework of meson Bethe-Salpeter equation together  with quark Dyson-Schwinger equation. Based on the zeroth approximation, all the elements, such as dressed quark propagator, dressed quark-gluon vertex, meson BSA and quark-anti-quark interaction kernel, are expanded up to the first order of $\epsilon$. Thereafter, the equation of the first order perturbative quark propagator is derived, a unique solution can be found due to its linearity. According to the perturbative theory of matrix, the first order correction of the eigenvalue of the BSE is derived. For the special case of pion, we rearrange the expansion of the $\lambda(\epsilon)$ so that the pion is massless at every order in the chiral limit, respecting the pion's Goldstone boson nature.

Employing Munczek-Nemirovsky model, we calculate the dressed propagator, $\pi$ and $\rho$ meson mass beyond RL truncation with our method. Our perturbatively obtained results are all in semi-quantitative agreement with the full solutions. Our method can therefore be used to give a quick estimate of meson mass beyond RL truncation. Meanwhile, it is a general method, which potentially allows the analysis of meson BSE with complicated interaction kernels if the dominant part are specified. We remind that although the RL truncation is dominant in ground state pseudoscalar and vector mesons, even the ground state baryon, in other cases it is not and the RL is not representative of QCD-like truncations. In that case treating the rival interaction terms perturbatively would allow a qualitative and preliminary estimate of the mass shift.

Finally, our method can be generalized to the baryon study, since the three-body bound state equation, i.e. the Faddeev equation, can also be converted into an eigenvalue problem of matrix. We note that beyond-RL truncation studies on baryon within a genuine three-body approach has been pioneered by Refs.~\cite{SANCHISALEPUZ2014151,PhysRevD.90.096001,SANCHISALEPUZ2015592}.

\acknowledgments
This work is supported in part by the National Natural Science Foundation of China (under Grant Nos. 11905107, 11905104), the National Natural Science Foundation of Jiangsu Province of China (under Grant No. BK20190721), Natural Science Foundation of the Jiangsu Higher Education Institutions of China (under Grant No. 19KJB140016), Nanjing University of Posts and Telecommunications Science Foundation (under grant No. NY129032), Innovation Program of Jiangsu Province.
\bibliography{references}
\end{document}